\documentstyle[prb,aps]{revtex}

\newcommand{\ie}{{\it i.e.}}
\newcommand{\eg}{{\it e.g.}}

\newcommand{\BR}[1]{\linebreak[0]#1\linebreak[0]}

\draft

\begin{document}

\title{
  Maximum velocity of a fluxon in a stack of coupled Josephson
  junctions
}

\author{E.~Goldobin\cite{FZJ}}
\address{
  Institute of Thin Film and Ion Technology,
  Research Center J\"ulich GmbH (FZJ),
  D-52425, J\"ulich, Germany\\
}

\author{B.~A.~Malomed\cite{TelAviv}}
\address{
  Department of Interdisciplinary Studies, Faculty of Engineering,
  Tel Aviv University, Tel Aviv 69978, Israel\\
}

\author{A.~V.~Ustinov}
\address{
  Physikalisches Institut III,
  Uni\-ver\-si\-t\"at Er\-lan\-gen-N\"urn\-berg,
  D-91058, Erlangen, Germany
}
\date{\today}


\maketitle

\begin{abstract}
  Dynamics of a fluxon in a stack of inductively coupled long
  Josephson junctions is studied analytically and numerically. We
  demonstrate that the fluxon has a maximum velocity, which does
  not necessarily coincide with any of the characteristic Josephson plasma wave velocities. The maximum fluxon velocity is found by means of numerical simulations of the quasi-infinite system. Using the variational approximation, we propose a simple analytical formula for the dependence of the fluxon's maximum velocity on the coupling constant and on the distribution of critical currents in different layers. This analysis yields rather precise results in the limit of small dissipation. The simulations also show that nonzero dissipation additionally stabilizes the fluxon.
\end{abstract}



\section{Introduction}

Experimental and theoretical studies of magnetic flux quanta (fluxons) in {\em stacks} of inductively coupled long Josephson junctions (LJJ's) have recently attracted a great deal of attention \cite {SBP,PUPS,CurLock:Cryogen92,Cherry1}. The interest to the stacks is stimulated both by the fact that the high-$T_{c}$ superconductors, on the atomic level, have a naturally layered structure that is tantamount to an intrinsic Josephson stack \cite{Intrinsic}, and by the development of the (Nb-Al-AlO$_{{\rm x}}$)$_{N}$-Nb low-$T_{c}$ technology \cite{TechSLJJ10}, which is demonstrated fabrication of artificial stacks of up to 28 LJJ's, with a parameter spread between them $<10\%$ \cite {Thyssen:PhD-Thesis}.

A fundamental characteristic of a {\em single} LJJ is its Swihart velocity ${ \bar{c}}_{0}$, {\it i.e.\ }, a minimum phase velocity of the electromagnetic (Josephson plasma) waves propagating in the superconducting micro-strip structure \cite {Swihart}. Simultaneously, ${\bar c}_{0}$ is the maximum velocity for fluxons that correspond to topological solitons of the sine-Gordon model describing LJJ. In a system of $N$ linearly coupled junctions, the dispersion curve has $N$ branches corresponding to different modes of the linear electromagnetic waves propagating in the system. Accordingly, there are $N$ split (different) Swihart velocities ${\bar{c}}_{n}^{(N)}$, $n=1,2,\ldots ,N$ (see, Refs.~\onlinecite{Kleiner2D,suk:94}) such that ${\bar{c}}_{n}^{(N)}<{\bar{c}}_{n+1}^{(N)}$. For example, in the simplest case of two coupled junctions, there are two Swihart velocities ${\bar c}_{1}^{(2)}\equiv{\bar c }_{-}$ and ${\bar c}_{2}^{(2)}\equiv{\bar c }_{+}$ (${\bar{c}}_{-}<{\bar{c}}_{0}<{\bar{c}}_{+}$), which correspond, respectively, to the system's in-phase and out-of-phase Josephson plasma wave eigenmodes. In the $N$-fold stack, we also use notation ${\bar{c}}_{-}\equiv\bar{c}_{1}^{(N)} $ and ${\bar{c}}_{+}\equiv\bar{c}_{N}^{(N)}$, which are the smallest and largest Swihart velocities.

An important issue is to study the conditions for the existence and
stability of a single-fluxon state in the stacked system. It is
implied that the fluxon is trapped in one junction and its screening
currents spread over neighboring junctions. The fluxon induces,
through the magnetic coupling, ``images'' in adjacent layers, so that
a full solution for the single fluxon state in the stack includes both
the core topological soliton in the central layer and its
non-topological images in the other layers. Such a fluxon state we denote as $[0|\ldots|0|1|0|\ldots|0]$.

As mentioned above, in a standard single-barrier LJJ a fully stable fluxon with a velocity exceeding the Swihart velocity cannot exist. It was first suggested\cite{image}, and recently demonstrated theoretically and experimentally\cite{Cherry1,Cherry2}, that a fluxon may nevertheless move in a multilayer system with a velocity which exceeds the minimum phase velocity of the plasma waves. It is important to find the maximum fluxon velocity $u_{\max }$ and its dependence on the parameters of the system. A possibility of having $u_{\max }>{\bar{c}}_{-}$ is especially interesting, as it implies steady motion of the fluxon at $u>{\bar c}_{-}$ with Cherenkov radiation tail of Josephson plasma waves behind it. This issue, which is of evident physical interest, is the main subject of the present work. The possible range of $u_{\max}$ was not investigated systematically in Refs.~\onlinecite{Cherry1,Cherry2}. The only prediction which has been made is that $u_{\max }>{\bar{c}}_{-}$ for asymmetric junctions in a two-fold stack (for the case when the fluxon is trapped in LJJ with lower $j_c$), and that $u_{\max }>{\bar{c}}_{-}$ always holds in $N$-fold stacks of identical junctions. The exact value of $u_{\max}$ has not been found until now.

We will discuss three cases which differ by the number $N$ of the
coupled LJJ's. These cases are $N=2$, $3$, and $\infty$. For $N=2$ and
$N=3$ we will consider a system of {\em asymmetric} LJJ's with
different critical currents $j_{c}$, in which the maximum velocity
is different depending on where the fluxon is
trapped. In the case $N=3$ and $N=\infty $ we will assume that the core topological soliton is placed in the
central junction which will be labeled by $0$, while LJJ's above and
below the central one will be labeled by $1,2,\ldots ,\infty $ and
$-1,-2,\ldots ,-\infty $, respectively.

In section \ref{Sec:Model} we formulate the model, section \ref {Sec:Simulation} displays results of full PDE numerical simulations of the asymmetric model for the cases $N=2$ and $N=3$. To choose an analytical form for the fitting function which predicts the dependence $u_{\max}$ on junction parameters, in section \ref{Sec:VA} we use the variational approximation (VA). Although VA does not produce very accurate quantitative results, it predicts reasonable functional dependence for $u_{\max}$. Section \ref {Sec:Conclusion} concludes the work and summarizes the obtained results for different $N$.

\section{The Model}

\label{Sec:Model}

A model for $N$-fold stack of long Josephson junctions is well known
\cite{SBP,N-foldModel}:
\begin{eqnarray}
\left( \phi _{n}\right) _{xx} &=&(\phi _{n})_{tt}+\sin \phi
_{n}+\gamma
+\alpha (\phi _{n})_{t}  \nonumber \\
&&-S\left[ (\phi _{n-1})_{tt}+\sin \phi _{n-1}+\alpha (\phi
_{n-1})_{t}+(\phi _{n+1})_{tt}+\sin \phi _{n+1}+\alpha (\phi
_{n+1})_{t}+2\gamma \right] \,,  \label{model}
\end{eqnarray}
where $\phi _{n}$ is the Josephson phase across the $n$-th LJJ,
$n=-N/2\ldots{}N/2$, the coordinate $x$ and time $t$ are measured in
units of the Josephson length $\lambda_{J}$ and inverse plasma
frequency $\omega _{p}^{-1}$ of single-layer LJJ, $S<0$ is a
dimensionless coupling parameter\cite{N-foldModel}, $\alpha$ is a
dissipative constant, and $\gamma$ is the density of the bias current
flowing through the stack. We consider the most natural case when the
bias current is the same in all layers.

The model (\ref{model}) pertains to a stack consisting of an infinite number of junctions, or to an exotic configuration, in which the stack is closed in a loop in $z$ direction ($\phi _{N+1}\equiv\phi_1$). In practice, the equations for the edge (top and bottom) junctions include only a half of the coupling terms corresponding to the neighboring LJJ's. Note also that the model implies that all the junctions are identical. In particular, the Swihart velocity of each uncoupled LJJ, is ${\bar c}_0\equiv 1$ in the notation adopted hereafter.

In the case of $N=2$, a relevant version of the model (\ref{model}),
which takes into account the difference in the critical currents of
the LJJ's, writes as \cite{Radio}:
\begin{eqnarray}
  \frac{\left( \phi _{0}\right) _{xx}}{1-S^{2}}-
  \left( \phi _{0}\right) _{tt}-\sin \phi _{0}-
  \frac{S\left( \phi _{1}\right) _{xx}}{1-S^{2}} &=&
  \alpha \left( \phi _{0}\right) _{t}+\gamma
  \ ; \label{Eq:2:phi0} \\
  \frac{\left( \phi _{1}\right) _{xx}}{1-S^{2}}-\left( \phi
  _{1}\right) _{tt}- \displaystyle\frac{\sin \phi
  _{1}}{J}-\frac{S\left( \phi _{0}\right) _{xx}}{ 1-S^{2}} &=&\alpha
  \left( \phi _{1}\right) _{t}+\gamma
  \ , \label{Eq:2:phi1}
\end{eqnarray}
where $J=j_{c0}/j_{c1}$ is the ratio of the critical currents of the
two junctions. When considering the model (\ref{Eq:2:phi0}) and
(\ref{Eq:2:phi1} ) below, we will place the fluxon into the LJJ whose
phase is denoted as $\phi _{0}$.

Discussing the case of $N=3$, we impose the symmetry condition $\phi
_{1}\equiv \phi _{-1}$, which is natural when the fluxon moves in the
middle layer. Thus, we can write Eqs.~(\ref{model}) in the form
\begin{eqnarray}
\frac{1}{1-2S^{2}}\left( \phi _{1}\right) _{xx}-\left( \phi
_{1}\right)
_{tt}-\frac{\sin \phi _{1}}{J}-\frac{S}{1-2S^{2}}\left( \phi
_{0}\right)
_{xx} &=&\gamma -\alpha \left( \phi _{1}\right) _{t}
\ ; \label{Eq:Sym3:pm1} \\
\frac{1}{1-2S^{2}}\left( \phi _{0}\right) _{xx}-\left( \phi
_{0}\right)
_{tt}-\sin \phi _{0}-\frac{2S}{1-2S^{2}}\left( \phi _{1}\right) _{xx}
&=&\gamma -\alpha \left( \phi _{0}\right) _{t}\quad .
\label{Eq:Sym3:0}
\end{eqnarray}
Note the factor of 2 in the last term of the left-hand side of Eq.~(\ref
{Eq:Sym3:0}).

For further theoretical treatment of the case $N=\infty $, we will
assume that the coupling parameter $S$ is small. This assumption will
allow us to write a Lagrangian corresponding to the dynamical
equations (\ref{model}), which is a key ingredient of VA. In the case
of small $S$, neglecting terms $ \sim S^{2}$ and smaller, one can
easily transform Eqs.~(\ref{model}) into a simplified form:
\begin{equation}
\left( \phi _{n}\right) _{xx}-\left( \phi _{n}\right) _{tt}-\sin \phi
_{n}-S
\left[ \left( \phi _{n-1}\right) _{xx}+\left( \phi _{n+1}\right)
_{xx}\right]
=\alpha (\phi _{n})_{t}-\gamma \quad ,  \label{simplemodel}
\end{equation}
which we will use for further analysis of the $N=\infty$ case.

A fluxon steadily moving at a constant velocity $u$ can be described
by a solution to the above equations (without the $\alpha $ and
$\gamma $ terms) which depends on the single variable $\xi =C(x-ut)$.
The constant $C$ is introduced for renormalization purposes and will
be different in the cases $ N=2$, $3$, and $\infty $. Substituting
\begin{equation}
\xi \equiv \sqrt{\frac{1-S^{2}}{-S}}(x-ut)  \label{Eq:xi2}
\end{equation}
into Eqs.~(\ref{Eq:2:phi0}) and (\ref{Eq:2:phi1}) and neglecting the
$\alpha $ and $\gamma $ terms (which must be kept if one aims to find
an equilibrium velocity determined by the balance between the losses
and bias current, that is not our objective in this work), we get:
\begin{eqnarray}
  \sigma^{(2)}\phi _{0}^{\prime \prime }+\phi _{1}^{\prime \prime }
  -\sin\phi_{0} &=&0
  \ , \label{Eq:2:ODEa} \\
  \sigma^{(2)}\phi _{1}^{\prime \prime }+\phi _{0}^{\prime \prime }
  -\displaystyle\frac{\sin \phi _{1}}{J} &=&0
  \ .  \label{Eq:2:ODEb}
\end{eqnarray}
The parameter $\sigma^{(2)}$ (the subscript $2$ implies that the
definition is adjusted to the case $N=2$), that will be used instead
of the velocity, is defined as:
\begin{equation}
\sigma^{(2)}\equiv \displaystyle\frac{1-\left( 1-S^{2}\right)
u^{2}}{-S}\quad
.  \label{Eq:sigma2}
\end{equation}

For the 3-fold stack, introduction of the traveling coordinate
\begin{equation}
\xi \equiv \sqrt{\frac{1-2S^{2}}{-S}}(x-ut)  \label{Eq:xi3}
\end{equation}
transforms Eqs.~(\ref{Eq:Sym3:pm1}) and (\ref{Eq:Sym3:0}) into
\begin{eqnarray}
\sigma^{(3)}\phi _{1}^{\prime \prime }+\phi _{0}^{\prime \prime
}-\frac{\sin
\phi _{1}}{J} &=&0\quad ,  \label{Eq:3:ODEa} \\
\sigma^{(3)}\phi _{0}^{\prime \prime }+2\phi _{1}^{\prime \prime }-\sin
\phi
_{0} &=&0\quad ,  \label{Eq:3:ODEb}
\end{eqnarray}
where
\begin{equation}
\sigma^{(3)}\equiv \frac{1-\left( 1-2S^{2}\right) u^{2}}{-S}\quad .
\label{Eq:sigma3}
\end{equation}
And, finally, for the case $N=\infty $ the substitution of
\begin{equation}
  \xi \equiv (x-ut)/\sqrt{-S}
  \label{Eq:xiN}
\end{equation}
into Eqs.~(\ref{simplemodel}) yields
\begin{equation}
  \sigma^{(\infty)}\phi _{n}^{\prime \prime }+\phi _{n-1}^{\prime \prime }+\phi _{n+1}^{\prime \prime }-\sin \phi _{n}=0
  ,  \label{Eq:N:ODE}
\end{equation}
where
\begin{equation}
  \sigma^{(\infty)}\equiv \frac{1-u^{2}}{-S}
  \ .  \label{Eq:sigmaN}
\end{equation}

Thus, from the mathematical viewpoint, the issue is to look for
solutions of Eqs.~(\ref{Eq:2:ODEa}) and (\ref{Eq:2:ODEb}), or
(\ref{Eq:3:ODEa}) and (\ref {Eq:3:ODEb}), or of
Eqs. (\ref{Eq:N:ODE}) that describe the stationary fluxon. The
eventual objective is finding the maximum velocity $u_{\max }$, or the
{\em minimum} value of the parameter $\sigma $, beyond which the
fluxon solution does not exist. The fluxon solution is defined by the
following boundary conditions: $\phi _{0}(-\infty )=0$, $\phi
_{0}(+\infty )=2\pi $, and $\phi _{n\neq 0}(\pm \infty )=0$.

To conclude this section, we recall that the set of the split Swihart
velocities can be found in an exact form from the linearized version
of Eqs.~(\ref{model}), setting there $\gamma =\alpha =0$. These
velocities are given
by the following expression \cite{Kleiner2D,suk:94}:
\begin{equation}
c_{n}^{(N)}=\frac{1}{\sqrt{1-2S\displaystyle\cos \left( \frac{\pi
n}{N+1}
\right) }},\quad n=1,2,\ldots ,N.  \label{c_n}
\end{equation}
It is also useful to find the values of the parameter $\sigma$
corresponding to the minimum velocity ${\bar c}_{-}$ and the maximum velocity  among all the velocities given by Eq. (\ref{c_n}). From Eqs.~(\ref{Eq:sigma2}),
(\ref{Eq:sigma3}) and (\ref{Eq:sigmaN}), using Eq. (\ref{c_n}), we
find
\begin{eqnarray}
\sigma^{(2)}({\bar{c}}_{\pm }) &=&\mp 1\quad , \\
\sigma^{(3)}({\bar{c}}_{\pm }) &=&\mp \sqrt{2}\quad , \\
\sigma^{(\infty)}({\bar{c}}_{\pm }) &=&\mp 2\quad ,
\end{eqnarray}
while zero velocity corresponds to $\sigma^{(2,3,\infty)}=-1/S$.

We can immediately find the values of $\sigma_{\min}$ for some
specially selected values of $J$. Let us set $\sigma^{(2)}=1$ in Eqs.
(\ref{Eq:2:ODEa}) and (\ref{Eq:2:ODEb}). This reduces Eqs.
(\ref{Eq:2:ODEa}) and (\ref {Eq:2:ODEb}) to a simple algebraic
equation:
\begin{equation}
\sin \phi _{0}=\frac{\sin \phi _{1}}{J}.
\end{equation}
The state with a fluxon only in the first junction ($\phi_1$) can be
realized only for $J<1$\cite{Cherry1,Cherry2}. This means that
\cite{Cherry2}
\begin{equation}
  \sigma_{\min }^{(2)}(J=1)=1
  \, . \label{Eq:limit_sigma2}
\end{equation}
In the same fashion, setting $\sigma^{(3)}=\sqrt{2}$ in Eqs. (\ref{Eq:3:ODEa}) and (\ref{Eq:3:ODEb}), we get
\begin{equation}
  \sin \phi _{0}=\frac{\sqrt{2}}{J}\sin \phi _{1}
  . 
\end{equation}
This means that
\begin{equation}
  \sigma_{\min }^{(3)}(J=\sqrt{2})=\sqrt{2}
  \, . \label{Eq:limit_sigma3}
\end{equation}

A similar relation can be obtained for any $N$-fold stack. Using
Eqs.~(\ref {simplemodel}), we get the following result. The value
$\sigma _{\min }^{(2N-1)}$ in the stack consisting of $(2N-1)$ LJJ's is given by the maximum eigenvalue of the $N\times N$ matrix,
\begin{equation}
  \left(
    \begin{array}{rrrrr}
      0 & -1 &  &  &  \\
      -1 & 0 & -1 &  &  \\
      & \ddots  & \ddots  & \ddots  &  \\
      &  & -1 & 0 & -1 \\
      &  &  & -2 & 0
    \end{array}
  \right)
  .  \label{Eq:Matrix}
\end{equation}
This matrix has the size $N\times{}N$ due to the symmetry which, as we suppose, is present when the fluxon moves in the middle layer. In this case $2N-1$ coupled equations can be reduced to $N$ equations in the same way as we reduced 3 equations for $N=3$ to only 2 independent equations. As a result of this reduction, the $N$-th equation, which describes the junction containing the fluxon and corresponds to the last row of matrix (\ref{Eq:Matrix}), contains the factor of 2.

The values of $\sigma _{\min }$ calculated for some $N$ are summarized
in Table~\ref{Tab:sigma_min}. One can see that as $N\rightarrow \infty
$, $ \sigma _{\min }\rightarrow 2$. Similar to the above
considerations, the single-fluxon state exists only if $J<\sigma
_{\min }$. Note that this is, actually, quite a noteworthy result, which
means that, for any $N>2$, the steady motion of the fluxon accompanied
by the emission of the Cherenkov radiation is possible at $J=1$, \ie, in the uniform stack of identical LJJ's.

Another simple case is obtained by setting $J=0$. In this case
Eq.~(\ref{Eq:2:ODEb}) or Eq.~(\ref{Eq:3:ODEa}) yield $\phi _{1}=0$,
hence, the remaining equation [(\ref{Eq:2:ODEa}) or (\ref{Eq:3:ODEb}),
respectively] is nothing else but the single sine-Gordon equation,
which has proper solitonic solution (giving the $[1|0]$ state) only
for positive $\sigma$. Therefore
\begin{equation}
  \sigma_{\min}(0)=0, \mbox{ for any } N
  \ . \label{Eq:limit_sigma0}
\end{equation}
Physically this means that at large disbalance of critical currents
the maximum velocity of fluxon is approaching $1\equiv{\bar c}_{0}$.

\section{Numerical Results}

\label{Sec:Simulation}

In the region $u>{\bar{c}}_{-}$, \ie{} at fluxon velocities larger than the lowest phase velocity of plasma waves, the phase dynamics in a stack is quite complex. Analytically, it can not be reduced just to looking for solutions of unperturbed equation in the form $\phi =\phi (x-ut)$. Therefore, direct numerical simulations are necessary to get an insight into the problem.

The PDE's (\ref{Eq:2:phi0}) and (\ref{Eq:2:phi1}) for $N=2$ or (\ref {Eq:Sym3:0}) and (\ref{Eq:Sym3:pm1}) for $N=3$ were solved numerically using an explicit method [expressing $\phi ^{A,B}(t+\Delta t)$ as a function of $ \phi ^{A,B}(t)$ and $\phi ^{A,B}(t-\Delta t)$], and treating $\phi _{xx}$ with a five-point, $\phi _{tt}$ with a three-point and $\phi _{t}$ with a two-point symmetric finite-difference scheme\cite{SBP,Radio}. Equations were supplemented by the periodic boundary conditions, $\phi _{0}(x+\ell )\equiv \phi _{0}(x)+2\pi $, and $\phi _{1}(x+\ell )\equiv \phi _{1}(x)$, with the period $\ell =200$, so that the fluxon moves in the quasi-infinite system. Numerical stability was checked by doubling the spatial and temporal discretization steps $\Delta x$ and $\Delta t$ and checking its influence on the fluxon profiles and current-voltage curves (IVC). The values used for the simulations were $\Delta x=0.025$ and $\Delta t=0.00625$. To calculate the voltage in each point of the IVC, the instant voltage was averaged over the progressively increasing time intervals and also averaged over the length of the system. The following convergence criterion was adopted: the difference between the voltages obtained by averaging over two successive time intervals must not exceed $\delta V=5\cdot10^{-4}$. When the average voltage corresponding to the current $\gamma $ is found, the current is increased by a small amount $\delta \gamma =0.001$ to calculate the voltages at the next point of the IVC. We use the phases (and their derivatives) attained in the previous point of the IVC as the initial conditions for the next point. By gradually increasing $\gamma $ and, thus, $u$, we encountered a maximum value $u_{\max }$ of $u$ above which the single-fluxon mode becomes unstable, and the system switches into a resistive state.

The simulations of IVC's were carried out for the damping values
$\alpha =0.02$, $0.04$, $0.1$, in order to understand the effect of
the dissipation on $u_{\max}$. In experiment, at low temperatures,
$\alpha$ can be about $0.01$ and less. It turns that the simulations
of the system with $\alpha <0.02$ incurs unaffordable computational
expenses. Therefore, we focus on higher values of $\alpha $ and then
will try to extrapolate $u_{\max}$ (if possible) to the ideal case
$\alpha =0$, which is of fundamental theoretical importance.

Here, we will only display the numerical results obtained for the cases $N=2$ and $N=3$. Examples of IVC's for $N=2$ and $3$ and different values of $J$ are shown in Fig.~\ref{Fig:2:SimIVC} and Fig.~\ref{Fig:3:SimIVC}. Strong effect of the ratio $J$ between the critical currents of the junctions on $u_{\max}$ can be learned from Fig.~\ref{Fig:2:SimIVC}. In the case $J=2 $, {\it i.e.\ }, when the fluxon is located in the junction with the {\em larger} critical current, the fluxon's maximum velocity is {\it smaller} than the lowest Swihart velocity ${\bar c}_{-}$. In addition, one can observe a back bending of the IVC close to its tip. The last point on back bent IVC marked as $[1|+1,-1]$ corresponds to the state when fluxon-antifluxon pair stretched out in the idle junction with lower $j_c$. In the case $J=0.5$, \ie, when the fluxon moves in the junction with the {\em smaller} critical current, the fluxon's maximum velocity is {\it larger} than ${\bar c}_{-}$. We note that the latter case is nontrivial and of the particular interest since, as it was already mentioned above, the fluxon can propagate stably, emitting the Cherenkov radiation.

Fig.~\ref{Fig:3:SimIVC} shows IVC's of the $[0|1|0]$ state for $J=1$ and different values of $\alpha$. As long as we consider 3-fold stack with $J<\sqrt{2}$, the IVC bends to the right into the Cherenkov region. The fluxon motion in such a case is very similar to the case $J<1$, $N=2$. Fig.~\ref{Fig:3:SimIVC} demonstrates that the damping not only changes the slope of the IVC at low velocity, but also affects $u_{\max}$.

The numerically found values $u_{\max}$ for different values of $J$ are transformed into the corresponding values of the parameter $\sigma _{\min }$ according to Eqs.~(\ref{Eq:sigma2}) and (\ref{Eq:sigma3}). The values of $\sigma _{\min }$ for $N=2$ and $N=3$ are plotted in Fig.~\ref {Fig:sigma_min(J)_4_N=2} and Fig.~\ref{Fig:sigma_min(J)_4_N=3}, respectively. Note, that the numerically obtained dependence $\sigma_{\min}(J)$ is in agreement with our analytical predictions given by Eqs.~(\ref{Eq:limit_sigma2}), (\ref{Eq:limit_sigma3}) and (\ref{Eq:limit_sigma0}). These three conditions work better for small $\alpha$ which is quite natural since they were derived from the unperturbed ($\alpha =\gamma =0$) equations. We remind that predictions (\ref{Eq:limit_sigma2}), (\ref{Eq:limit_sigma3}) and (\ref{Eq:limit_sigma0}) are strict results and must be considered exact in the framework of inductive-coupling model. We found it interesting that the dependence of the maximum current $\gamma_{\max}=\gamma(u_{\max})$ on $J$ is almost linear with $\gamma_{\max}/J=0.476$, 0.531, 0.623 for $\alpha=0.02$, 0.04, 0.10, respectively.

In the following section we suggest a functional dependence for $\sigma_{\min}(u)$ and will compare this dependence with numerical results presented in Fig.~\ref {Fig:sigma_min(J)_4_N=2} and Fig.~\ref{Fig:sigma_min(J)_4_N=3}.

\section{The Variational Approximation}
\label{Sec:VA}

The purpose of this section is to find an analytical dependence which describes the simulation data obtained in the previous section. We will present an analysis using VA only for $N=2$, but other cases can be considered similarly.

VA is based on the fundamental fact that Eqs.~(\ref{Eq:2:ODEa}) and
(\ref {Eq:2:ODEb}) admit the variational representation with the
Lagrangian $L=\int_{-\infty }^{+\infty }{\cal L}_2\,d\xi$, where the
{\it Lagrangian density} corresponding to Eqs.~(\ref{Eq:2:ODEa}) and
(\ref{Eq:2:ODEb}) is
\begin{equation}
  {\cal L}_{2}=\frac{1}{2}\sigma \left( \phi _{0}^{\prime }\right)
  ^{2}+\frac{1 }{2}\sigma \left( \phi _{1}^{\prime }\right) ^{2}+\phi
  _{0}^{\prime }\phi _{1}^{\prime }+\left( 1-\cos \phi _{0}\right)
  +\frac{1}{J}\left( 1-\cos \phi _{1}\right)
  \quad .  \label{Eq:2:L-density}
\end{equation}
A crucial step in the application of VA is to adopt an {\em ansatz},
{\ie}, a trial form of the solution. The ansatz is then to be inserted
into the corresponding Lagrangian, and the integration over $\xi$
given by Eq.~(\ref {Eq:xi2}) must be performed explicitly. This will
produce an {\it effective Lagrangian} as a function of free parameters
that the ansatz contains. Finally, the values of this parameters will
be found from the condition that they must realize an extremum of the
effective Lagrangian.

Thus, it is necessary to select a tractable ansatz (that admits an
analytical calculation of the integrals in the expression for the
Lagrangian), which satisfies the above boundary conditions for the
fluxon. Here, both when selecting the ansatz and considering the
boundary conditions, we ignore the above-mentioned nonvanishing Cherenkov oscillatory tails (note that the tail is formally infinitely long in the ideal model with $\alpha =\gamma =0$, while in the damped system the tail decays exponentially far from the fluxon's ``body''). The simplest and, in fact, the only practical ansatz is
\begin{eqnarray}
  \phi _{0}(\xi ) &=&4\arctan \exp (\lambda \xi )
  \ , \label{0ansatz} \\
  \phi _{n\neq 0} &=&
  B\frac{\sinh (\lambda \xi )}{\cosh ^{2}(\lambda \xi )}
  \ . \label{nansatz}
\end{eqnarray}
Here, $\lambda$ and $B$ are free real parameters. Note that $\lambda $ can always be defined positive, while $B$ may be positive as well as negative. The expression (\ref{nansatz}) was chosen in such a way that it describes the ``image'' profile qualitatively well in comparison with simulation and analytical results\cite{image}.

Still, the effective Lagrangian cannot be immediately calculated with
the above ansatz. A further necessary simplification is to assume that
the amplitude $B$ of the fluxon's images in the noncore layers is
sufficiently small, so that the nonlinear term $\sin \phi _{1}$ in
Eq.~(\ref{Eq:2:ODEb}) may be linearized. In other words, the term
$\left( 1-\cos \phi _{1}\right) $ in the Lagrangian density
(\ref{Eq:2:L-density}) is to be replaced by $\frac{ 1}{2}\phi
_{1}^{2}$.

The final expression for the effective Lagrangian is
\begin{equation}
  L=4\sigma \lambda +\frac{4}{\lambda}+\frac{7}{15}\sigma B^{2}\lambda
  +\frac{ 1}{3}\frac{B^{2}}{J\lambda}+\frac{4}{3}B\lambda
  \ .  \label{L2}
\end{equation}
The variation of (\ref{L2}) in $B$ leads to an equation that allows us to eliminate
$B$, {\it i.e.,\ }
\begin{equation}
B=-\frac{10J\lambda ^{2}}{\left(
5+7J\sigma \lambda ^{2}\right) }
\ . \label{Eq:2:B}
\end{equation}
The remaining algebraic equation for the fluxon's inverse width
$\lambda $ is
\begin{equation} 7J^{2}\sigma \left( 21\sigma
^{2}-5\right) \lambda ^{6}-3J\left[ 25-7\sigma ^{2}\left( 10-7J\right)
\right] \lambda ^{4}-15\sigma \left( 14J-5\right) \lambda ^{2}-75=0 \,
.  \label{Eq:cumbersome}
\end{equation}
This equation has one or
three positive roots for $\lambda ^{2}$. The first two solutions exist
and disappear simultaneously at some value of $\sigma $. The third
solution disappears diverging at $\sigma _{\min }=\sqrt{5/21}$. This
can be seen from the form of the coefficient in front of the term
$\sim \lambda ^{6}$. We drop the third solution from the consideration
since it gives a constant $\sigma _{\min }(J)$ which is unphysical due
to Eqs. (\ref {Eq:limit_sigma2}) and (\ref{Eq:limit_sigma0}).

To find the minimum value of $\sigma $ at which two smallest positive
roots disappear, we write two conditions: the function $f(\lambda )$
(\ref {Eq:cumbersome}) touches $\lambda $-axis and its derivative
$f^{\prime }(\lambda )$ vanishes. These two equations determine $\sigma
_{\min }$ and $ \lambda (\sigma _{\min })$ for given value of $J$.
Eliminating, consecutively, the terms $\sim \lambda ^{6}$, $\lambda
^{4}$ and $\lambda ^{2}$ yields an equation which determines the
dependence of $\sigma _{\min }$ on $J$:
\begin{equation}
  56\left( 7J+5\right) ^{3}\sigma _{\min }^{4}-\left(
  3675J^{2}+36750J+1875\right) \sigma _{\min }^{2}+7500J=0
  \, , \label{Eq:2:sigma(J)}
\end{equation}
A solution to this equation is
\begin{equation}
  \sigma _{\min }^{2}=\frac{36750J+1875+3675J^{2}\pm
  125\sqrt{15}\sqrt{ (15+7J)\left( 1-7J\right) ^{3}}}{112(5+7J)^{3}}
  \, . \label{Eq:2:ExactSigma}
\end{equation}
One can see that this solution exists only for $J<1/7$. For $J>1/7$,
Eqs.~( \ref{Eq:cumbersome}) has only one (third) root, which is
unphysical.

To relax the latter limitation, we can make use of the fact that, as
it was said above, the asymmetry parameter $J$ always takes values
$J<1$. We will make use of this, treating $J$ as a {\em small}
parameter, which is also justified by the fact that the stack with a
sufficiently strong asymmetry might be of special physical interest.
With regard to this, VA leads to a simple expression for $\sigma
_{\min }$ (we just expand Eq. (\ref{Eq:2:ExactSigma}) in a Taylor
series around $J=0$):
\begin{equation}
  \sigma _{\min }\approx 2\sqrt{J}
  \,.\label{simplest}
\end{equation}
The essential feature in Eq.~(\ref{simplest}) is a square root
dependence. To comply with the strict analytical result for the unperturbed case, see Eqs. (\ref{Eq:limit_sigma2}) and (\ref{Eq:limit_sigma0}), we adopt
\begin{equation}
  \sigma_{\min }^{(2)}\approx \sqrt{J}
  ,\label{Eq:sigma_2(J)}
\end{equation}
For $N=3$, a similar dependence is found to be
\begin{equation}
  \sigma_{\min }^{(3)}\approx \sqrt{\sqrt{2}J}
  .  \label{Eq:sigma_3(J)}
\end{equation}
as an approximation. The bold solid lines corresponding to these
approximations are shown in Fig.~\ref{Fig:sigma_min(J)_4_N=2} and Fig.~\ref{Fig:sigma_min(J)_4_N=3}.

Apparently, the smaller the dissipation, the better the approximations
(\ref {Eq:sigma_2(J)}) and (\ref{Eq:sigma_3(J)}) work. In the presence
of the dissipation, the fluxons can exist at velocities somewhat
larger than in the idealized model. The latter circumstance is quite
natural, as in the single Josephson junction the onset of the fluxon's
instability past the Swihart velocity is also delayed by the
dissipation \cite{BL}.

We can also obtain an analytical approximation for $N\rightarrow
\infty$. In this case,
\begin{equation}
  \sigma _{\min } \approx \sqrt{2J}
  \,.  \label{Eq:sigma-min}
\end{equation}
The stack with the uniform critical current distribution, therefore,
has $\sigma _{\min }=\sqrt{2}$. The Cherenkov radiation will appear in
the range $\sqrt{2}<\sigma _{\min }<2$.

It is convenient to present the results of our calculations as the plot $u_{\max}(|S|)$. Such a plot for $N=3$ and $\sigma^{(3)}=2^{1/4}$ (uniform stack) is shown in Fig.~\ref{Fig:u-max(S)}. The region where the fluxon moves faster than ${\bar c}_{-}$ is shaded. It is the domain of existence of the Cherenkov radiation. The range of $S$ in Fig.~\ref{Fig:u-max(S)} corresponds to the maximum value of the coupling parameter $|S|_{\max}$ in 3-fold stack which is equal to $1/\sqrt{2}$.

\section{Conclusion}
\label{Sec:Conclusion}

We have demonstrated that a single fluxon moving in the stack of Josephson junctions has the maximum velocity which does not necessarily coincide with one of the Swihart velocities of the system. The dependence $u_{\max }(J)$ was studied numerically for $ N=2$ and $3$. An analytical approximation for this dependence was put forward on the basis of the variational approximation in the limit of small dissipation and is given by Eqs.~(\ref{Eq:sigma_2(J)}) and (\ref {Eq:sigma_3(J)}). Our results show that $u_{\max }>{\bar{c}}_{-}$ for $J<1$, in the case $N=2$, and for $J<\sqrt{2}$, in the case $N=3$. This leads to Cherenkov radiation of plasma waves by a fluxon in the velocity range ${\bar{ c}}_{-}<u<u_{\max }$. Simulations also show that the damping stabilizes the fluxon motion at higher velocities, {\it i.e.\ }, $u_{\max }(\alpha>0)>u_{\max }(\alpha=0)$. In uniform stack with large $N$ (\eg{} intrinsic Josephson stacks in crystals of high-$T_c$ superconductors), $u_{\max }$ exceeds ${\bar{c}} _{-}$ and, therefore, a single fluxon always generates Cherenkov radiation. Experiments with high-$T_c$ stacks \cite{Cherry:HTS} show that flux-flow branches do not have vanishing differential resistance $R_{d}$ close to the top of the flux-flow step as in the single LJJ but bend towards higher voltages, in agreement with our results. From the theoretical point of view, the absence of the ``relativistic'' singularity at $u={\bar{c}}_{-}$ is a result of the lack of the Lorentz invariance in the coupled sine-Gordon equations.

\acknowledgments

B.A. Malomed appreciates hospitality of the Department of Physics at
the University of Erlangen-N\"urnberg. This work is supported by a grant no. G0464-247.07/95 from German-Israeli Foundation and, in part, by Deutsche Forschungsgemindschaft (DFG).


\section*{Figure Captions}

\begin{figure}[tbp]
  \caption{
    The IVC's $u(\protect\gamma)$ of a 2-fold stack with a single fluxon trapped in one LJJ (the state $[1|0]$) for three different values of $J=0.5$, $1$, and $2$. The branches for $J=0.5$ and $J=1$ terminate at a value $u_{\max}$ at which the fluxon becomes unstable and the system switches to high voltages. The inset is a blowup of the rectangular area on the main diagram. The simulations were performed for $\alpha=0.1$ and demonstrate the dependence of $u_{\max}$ on $J$.
  }
  \label{Fig:2:SimIVC}
\end{figure}
\begin{figure}[tbp]
  \caption{
    The IVC's of the 3-fold stack with a fluxon in the central layer for three different values of $\protect\alpha=0.02$, $0.04$, and $0.1$. The simulations performed for $J=1$ demonstrate the dependence of $u_{\max}$ on $\alpha$.
  }
  \label{Fig:3:SimIVC}
\end{figure}
\begin{figure}[tbp]
  \caption{
    The dependence of $\protect\sigma^{(2)}_{\min}$ on $J$ for three
    different values of $\protect\alpha=0.02$ (circles), $0.04$
    (squares), and $0.10$ (diamonds). The bold solid line shows the
    analytical dependence (\ref{Eq:sigma_2(J)}) produced by the
    variational approximation.
  }
  \label{Fig:sigma_min(J)_4_N=2}
\end{figure}
\begin{figure}[tbp]
  \caption{
    The dependence of $\protect\sigma^{(3)}_{\min}$ on $J$ for three
    different values of $\protect\alpha=0.02$ (circles), $0.04$
    (squares), and $0.10$ (diamonds). The bold solid line shows the analytical prediction (\ref{Eq:sigma_3(J)}) based on the variational
    approximation.
  }
  \label{Fig:sigma_min(J)_4_N=3}
\end{figure}
\begin{figure}[tbp]
  \caption{
    The dependence of maximum velocity $u_{\max}$ and Swihart velocities ${\bar c}_{\pm}$ on the coupling strength $|S|$ for three coupled junctions at $J=1$. The shaded area shows the domain of the Cherenkov radiation.
  }
  \label{Fig:u-max(S)}
\end{figure}

\begin{table}[tbp]
  \begin{tabular}{c|cccccccc}
    $N$ & 3 & 5 & 7 & 9 & 13 & 19 & 39 & 99 \\ \hline
    $\sigma_{\min}$ &1.414&1.732&1.848&1.902&1.950&1.975&1.994&1.999
  \end{tabular}
  \caption{
    The values of $\protect\sigma_{\min}$ for different numbers $N$
    of junctions in the stack. As $N\to\infty$,
    $\protect\sigma_{\min}\to2$.
  }
  \label{Tab:sigma_min}
\end{table}


\begin{references}

\bibitem[{*}]{FZJ}
 e-mail: e.goldobin@fz-juelich.de,
 homepage: http:\BR{//}www\BR{.}geocities\BR{.}com\BR{/}e\_goldobin

\bibitem[\dag ]{TelAviv}
  e-mail: malomed@ayalon.eng.tau.ac.il


\bibitem{SBP}
  S.~Sakai, P.~Bodin, and N.~F.~Pedersen,
  J.~Appl. Phys. {\bf 73}, 2411 (1993).

\bibitem{PUPS}
  A.~Petraglia, A.~V.~Ustinov, N.~F.~Pedersen, S.~Sakai,
  J.~Appl.~Phys. {\bf 77}, 1171 (1995).

\bibitem{CurLock:Cryogen92}
  I.~P.~Nevirkovets, H.~Kohlstedt, C.~Heiden,
  ICEC Suppl., Cryogenics {\bf 32}, 583 (1992).

\bibitem{Cherry1}
  E.~Goldobin, A.~Wallraff, N.~Thyssen, and A.~V.~Ustinov,
  Phys. Rev. B {\bf 57}, 130 (1998).

\bibitem{Intrinsic}
  R.~Kleiner, F.~Steinmeyer, G.~Kunkel, and P.~M\"uller,
  Phys. Rev. Lett. {\bf 68}, 2394 (1992);\\
  R.~Kleiner and P.~M\"uller,
  Phys. Rev.~B {\bf 49}, 1327 (1994).

\bibitem{TechSLJJ10}
  H.~Kohlstedt, G.~Hallmanns, I.~P.~Nevirkovets,
  D.~Guggi, and C.~Heiden,
  IEEE Trans. Appl. Supercond. {\bf 3}, 2197 (1993).

\bibitem{Thyssen:PhD-Thesis}
  N.~Thyssen, Ph.D. thesis,
  Universit\"{a}t Erlangen-N\"urnberg (1999).

\bibitem{Swihart}
  J.~C.~Swihart,
  J.~Appl. Phys. {\bf 32}, 461 (1961).

\bibitem{Kleiner2D}
  R.~Kleiner,
  Phys. Rev. B {\bf 50}, 6919 (1994).

\bibitem{suk:94}
  S.~Sakai, A.~V.~Ustinov, H.~Kohlstedt,
  A.~Petraglia, and N.~F.~Pedersen,
  Phys. Rev.~B {\bf 50}, 12905 (1994).

\bibitem{image}
  Yu.~S.~Kivshar, B.~A.~Malomed,
  Phys. Rev.~B {\bf 37}, 9325 (1988).

\bibitem{Cherry2}
  E.~Goldobin, A.~Wallraff, and A.~V.~Ustinov,
  cond-mat/9910234, submitted to J. Low Temp. Phys (Sept 1998).

\bibitem{N-foldModel}
  S.~Sakai, A.~V.~Ustinov, N.~Thyssen and H.~Kohlstedt,
  Phys. Rev. B, {\bf 58}, 5777 (1998).

\bibitem{Radio}
  A.~Wallraff, E.~Goldobin, A.~V.~Ustinov,
  J.~Appl. Phys. {\bf 80}, 6523 (1996).

\bibitem{BL}
  S.~E.~Burkov, A.~E.~Lifshits,
  Wave Motion {\bf 5}, 197 (1983)

\bibitem{Cherry:HTS}
  G.~Hechtfischer, R.~Kleiner, A.~V.~Ustinov, and P.~M\"uller,
  Phys. Rev. Lett. {\bf 79}, 1365 (1997).


\end{references}
\end{document}